\documentclass[twocolumn]{aastex631}

\begin{document}

\title{AT2022sxl: A Candidate Repeating Tidal Disruption Event in Possible Association with Two High-Energy Neutrino Events}

\author{Shunhao Ji}
\affiliation{Department of Astronomy, School of Physics and Astronomy, Key Laboratory of Astroparticle Physics of Yunnan Province, Yunnan University, Kunming 650091, China; jishunhao@mail.ynu.edu.cn, wangzx20@ynu.edu.cn}

\author[0000-0003-1984-3852]{Zhongxiang Wang}
\affiliation{Department of Astronomy, School of Physics and Astronomy, Key Laboratory of Astroparticle Physics of Yunnan Province, Yunnan University, Kunming 650091, China; jishunhao@mail.ynu.edu.cn, wangzx20@ynu.edu.cn}
\affiliation{Purple Mountain Observatory, Chinese Academy of Sciences, Nanjing 210034, China}

\author{Litao Zhu}
\affiliation{Department of Astronomy, School of Physics and Astronomy, Key Laboratory of Astroparticle Physics of Yunnan Province, Yunnan University, Kunming 650091, China; jishunhao@mail.ynu.edu.cn, wangzx20@ynu.edu.cn}

\author{Stefan Geier}
\affiliation{GRANTECAN: Cuesta de San Jos\'{e} s/n, 38712 Bre{\~n}a Baja, La Palma, Spain}
\affiliation{Instituto de Astrof\'{\i}sica de Canarias, V\'{\i}a   L\'actea, 38205 La Laguna, Tenerife, Spain}

\author[0000-0002-9331-4388]{Alok C. Gupta}
\affiliation{Aryabhatta Research Institute of Observational Sciences (ARIES), Manora Peak, Nainital-263001, India}

\begin{abstract}

	We report a candidate repeating tidal disruption event (TDE),
	AT2022sxl, found from 
	large-field optical survey data.  Two flares with a separation time
	of $\sim$7.2\,yr between the two optical peaks are observed. 
	Related mid-infrared (MIR) flares, with delay times of $\sim$200\,day
	are also seen. We analyze two optical spectra of the TDE source, one
	near the optical peak of the second flare from the Transient Name
	Server and one at the quiescent flux level after the second flare.
	The latter was taken by us with the 10.4-m Gran Telescopio 
	Canarias. Comparing the features of the two spectra, we identify
	that the host is likely a composite galaxy at redshift 0.23 and
	the TDE event, probably an H+He type,  mainly powered broad 
	components in the emission lines of H$\alpha$, H$\beta$, and 
	He~I $\lambda$5876. More interestingly, we find that two Bronze-type 
	neutrino events, detected by the IceCube neutrino observatory, match
	the TDE in position and the second flare, especially the delayed MIR 
	flare, in time. We discuss the MIR luminosity properties of 
	the currently
	reported (candidate) neutrino-emitting TDEs and suggest that
	luminous MIR emission is a prerequisite for neutrino 
	production in TDEs.
\end{abstract}


\section{Introduction}

Tidal disruption events (TDEs) are one type of transient phenomena mostly
occurring around central supermassive black holes (SMBHs) of galaxies:
when a star approaches such an SMBH, it can be under too
large tidal forces, exerted by the latter, to keep itself gravitationally 
bounded and thus be disrupted \citep{ree88}. While TDEs are considered to
be rare, more than 100 of them have been discovered
(see, e.g., \citealt{gez21, van+21, ham+23, yao+23}), thanks to 
large-field optical transient surveys carried out in recent years,
in particular the Zwicky Transient Facility \citep{Bellm+19}. TDEs reveal 
the dormant SMBHs in galaxies and their transient accretion activities.
TDE Studies provide a powerful tool for probing properties of
SMBHs and related physical processes (e.g., \citealt{gez21}).

One recent progress in studies of TDEs is the discovery of a few
(candidate) repeating TDEs (rTDEs). The optical rTDEs were reported in
\citet{pay+21}, \citet{wev+23}, \citet{yao+23}, \citet{lin+24}, \citet{ver+24},\
\citet{sun+24}, and \citet{sun+25}.
In $\sim$0.3--10 years, repeating TDE flares were seen again in the same
galaxy or active galactic nucleus (AGN; see \citealt{sun+25} and references
therein). Such rTDEs are discussed to be likely caused by the partial
disruption of a star by an SMBH \citep{pay+21}. This star, with an eccentric
orbit around the SMBH, will power a TDE flare when it passes the pericenter
of the orbit. Thus, multiple flares may be seen from a rTDE case 
\citep{mac+13, pay+21}. However, the identification of true repeating partial
TDEs can be challenging, which requires clear spectroscopic evidence
(see detailed discussion in \citealt{lin+24}).

\begin{figure*}
    \centering
    \includegraphics[width=0.8\linewidth]{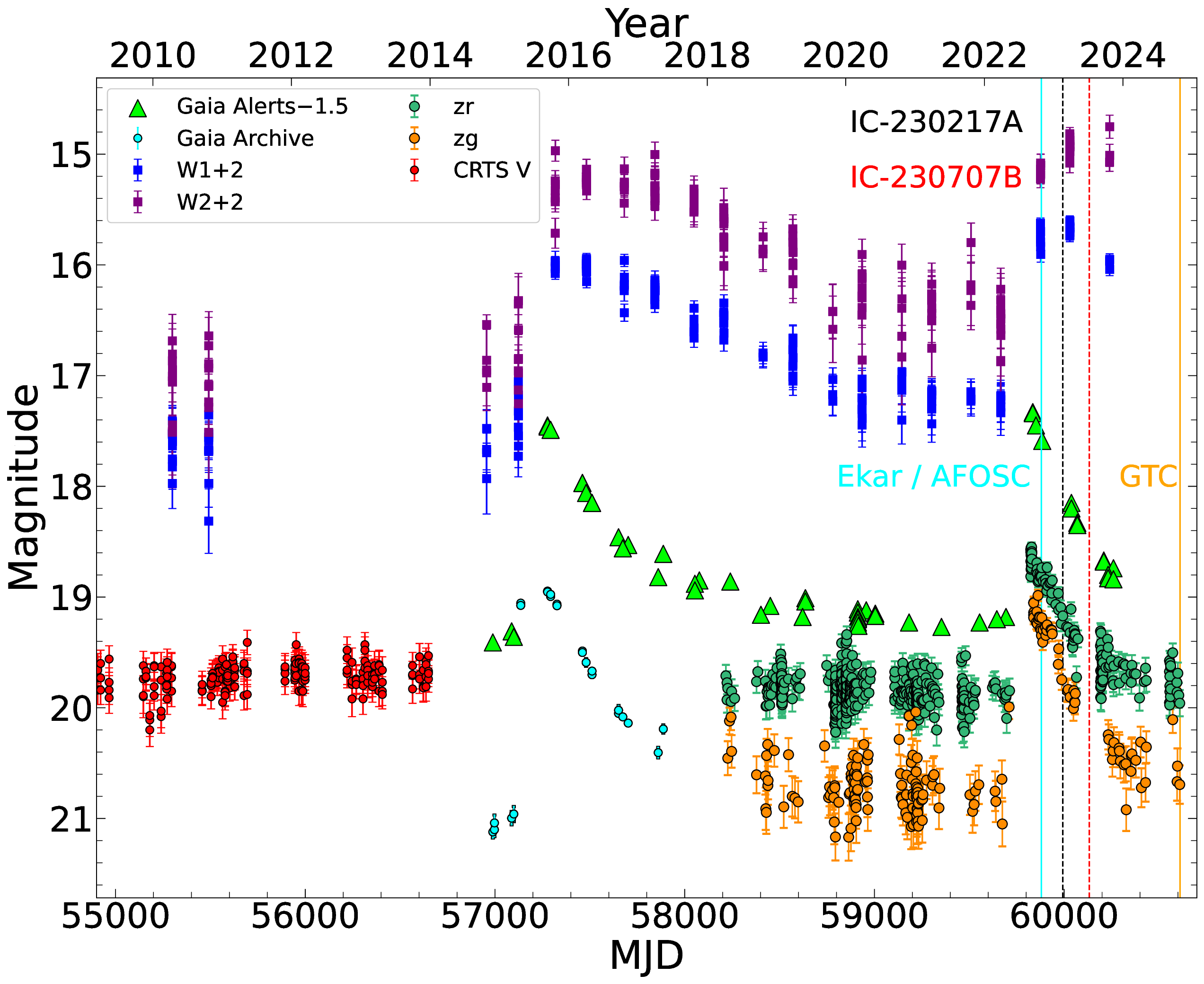}
    \caption{Optical and MIR light curves of AT2022sxl; for clarity, 
	the Gaia alerts data are upshifted by 1.5\,mag and the WISE MIR data are
	downshifted by 2\,mag. Two flares are clearly visible. The times
	of the TNS and GTC spectroscopic observations are marked with cyan
	and yellow solid lines, respectively. The two neutrino arrival
	times are marked with black and red dashed lines, respectively.}
    \label{fig:multi}
\end{figure*}

Another progress in studies of TDEs is their possible association with 
high-energy (HE) neutrinos. Since 2013, the IceCube neutrino observatory, built
at the South pole, has been detecting the HE neutrinos that are likely of an
astrophysical origin \citep{aar+13}. While most of the reported neutrino
events, for example 126 Gold alerts (defined by a quantity `signalness') 
in the IceCube Event Catalog of Alert Tracks (IceCat-1; \citealt{abb+23}), do
not have an identified origin, several types of objects have been proven to be
able to emit neutrinos. They are the blazar jets \citep{txs0506a,txs0506b}, 
nearby AGNs
\citep{ngc1068,abb+25}, and the Galactic plane \citep{gal+23}. In addition to
them, TDEs have also been considered as the candidate neutrino emitters because
several of them have been found to be in positional and temporal coincidence
with HE neutrino events  
\citep{ste+21, reu+22, jia+23, van+24, ywl24, li+24}.

In this paper, we report a likely TDE event, AT2022sxl, found from 
analyzing 
the archival photometric data at multi-bands. In order to check its
spectral variations, a spectroscopic observation of the source was 
conducted by us in 2024. In addition, 
a TDE-like flare previously occuring in $\sim$2016--2019 was revealed by the
multi-band data, thus suggesting a candidate rTDE case.
Very interestingly, we found that this source was in 
positional and temporal coincidence with two IceCube neutrino events.
Throughout the paper, we used the following cosmological parameters,
$H_0$ = 67.7 km s$^{-1}$ Mpc$^{-1}$, $\Omega_m$ = 0.32, and
$\Omega_\Lambda$ = 0.68 \citep{Planck18}.

\begin{figure*}
    \centering
    \includegraphics[width=0.48\linewidth]{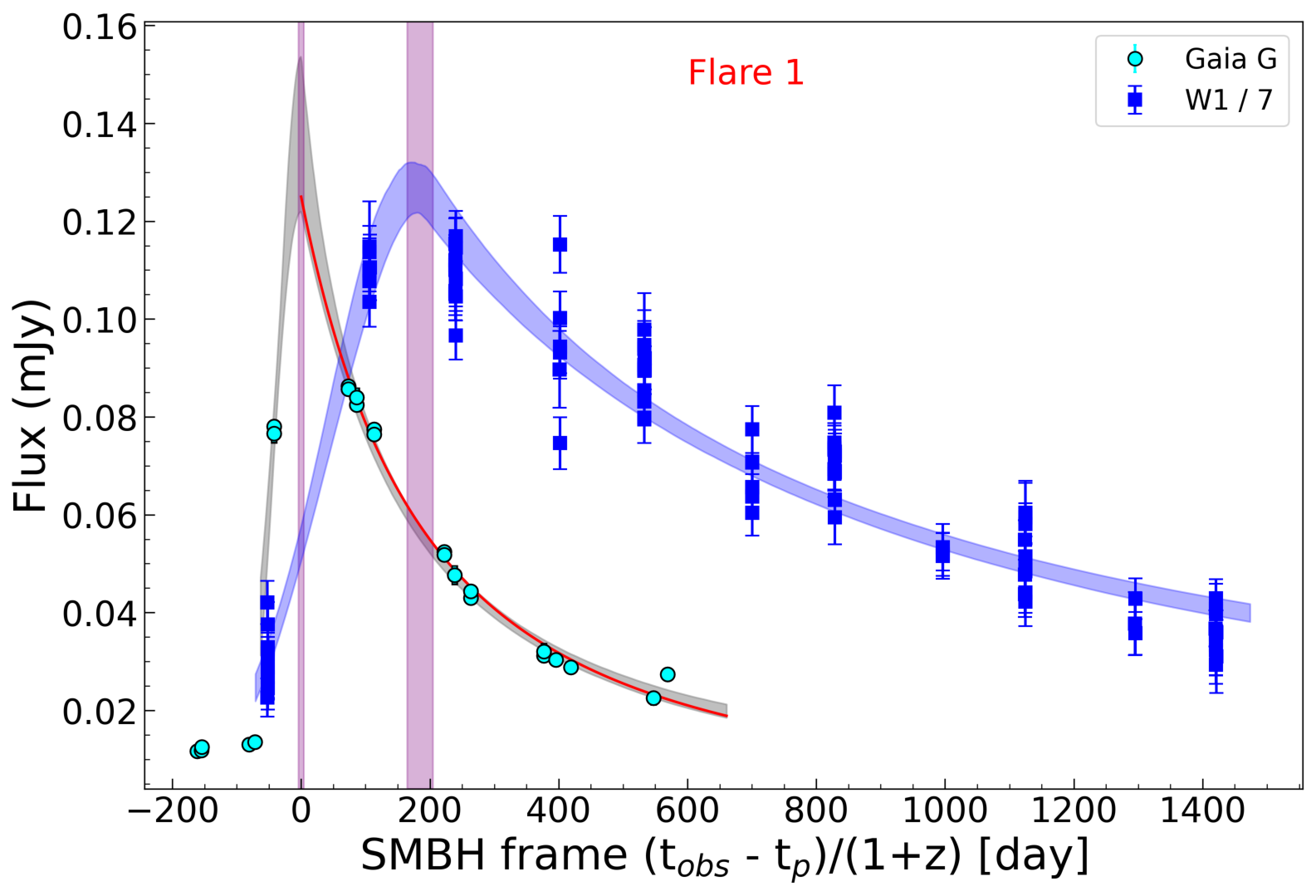}
    \includegraphics[width=0.49\linewidth]{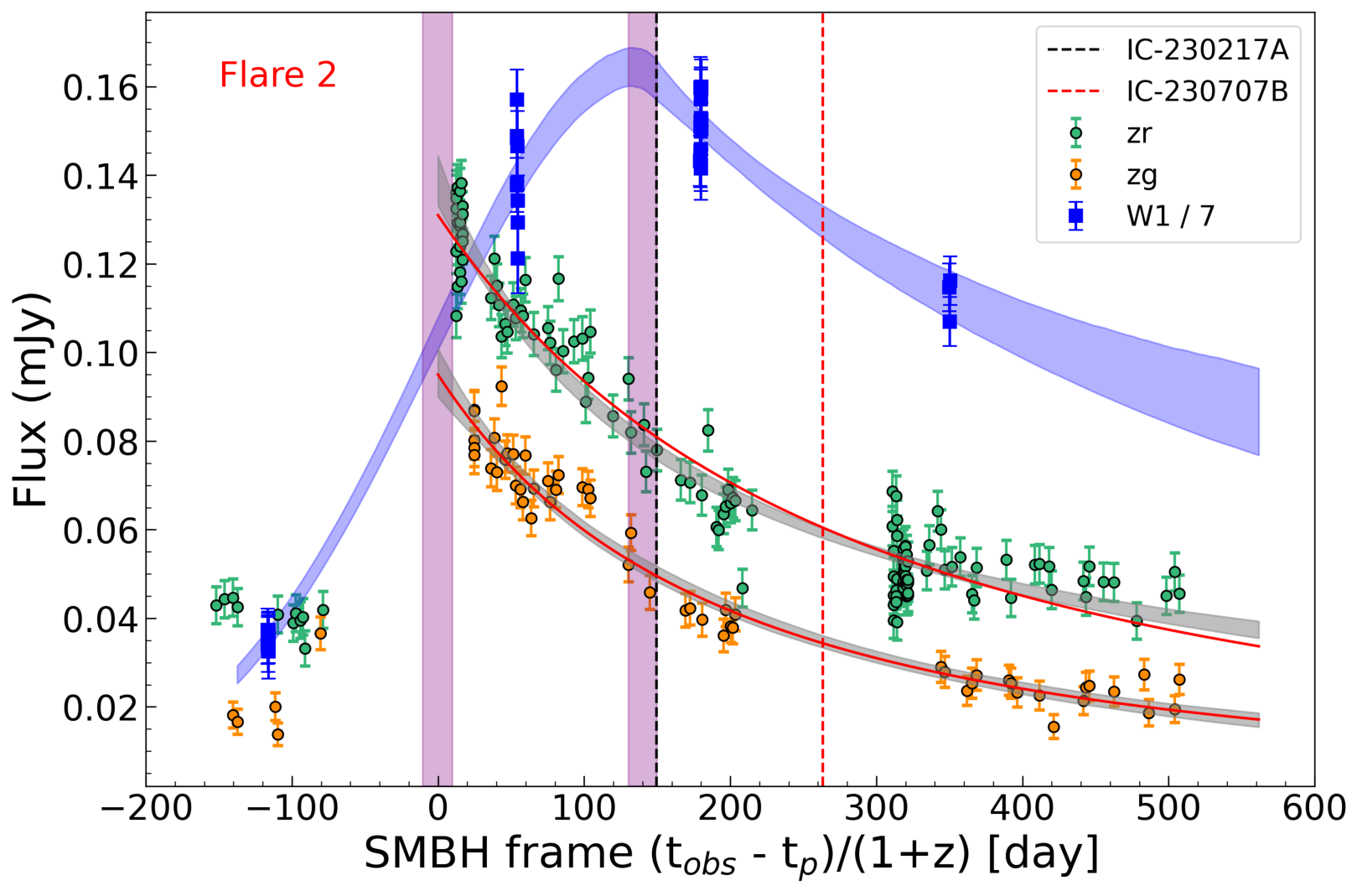}
    \caption{Fits to the light curves with the typical TDE model. In order to
	show the MIR ones in one panel for comparison, their flux densities 
	are divided by 7.
	The peak time ranges are marked by purple regions, while the optical
	ones are from the MOSFiT results. For flare 2 ({\it right} panel),
	the arrival times of the two neutrinos are marked by black and red
	dashed lines.}
    \label{fig:lc_mcmc}
\end{figure*}

\section{Data Analysis and Results}\label{ana}

\subsection{Photometric Data}

\subsubsection{Optical and mid-infrared data} \label{opt}

Optical light-curve data were obtained from the Zwicky Transient 
Facility (ZTF; \citealt{Bellm+19}), Catalina Real-Time Transient Survey 
(CRTS; \citealt{Drake+09}), and the European Space Agency (ESA) mission 
Gaia \citep{gaia16}. The bands of the data are ZTF $zg$ and $zr$, CRTS $V$,
and Gaia $G$. To obtain good-quality light curves from ZTF, we selected 
magnitude data points by requiring $catflags = 0$ and $chi < 4$. The Gaia 
$G$-band photometric data were obtained form Gaia DR3 Archive \citep{gaiaDR3} 
and Gaia Photometric Science Alerts\footnote{\url{http://gsaweb.ast.cam.ac.uk/alerts/alert/Gaia22dws/}}, and the fluxes in units of photo-electrons~s$^{-1}$ 
were converted to the fluxes in units of Jy and magnitudes by multiplying 
with the conversion factor given in the Gaia DR3 documentation. 
In addition, the mid-infrared (MIR) light-curve data were obtained from 
the AllWISE Multiepoch Photometry Database and the NEOWISE Single-exposure 
Source Database (\citealt{Mainzer+11};\citealt{neowise}), where the bands 
are WISE W1 (3.4\,$\mu$m) and W2 (4.6\,$\mu$m).

\subsubsection{Light-curve fitting analysis} \label{sec:opt}

The optical and MIR light curves of AT2022sxl are shown in 
Figure \ref{fig:multi}. As can be seen, two significant optical flares, 
the first peaking at $\sim$MJD~57200 (flare~1) and the second peaking 
at $\sim$MJD~59800 (flare~2), are observed. The two flares both exhibit 
the characteristics of having a fast rise and a slow decline. Probably because
the rises were so fast that few data points caught the rise phases. 
Accompanying 
the optical flares, the delayed MIR flares are also observed. In the catalog
Flaires \citep{Necker+25}, the first MIR flare was classified 
as a dust-echo-like one, which may be interpreted as an extreme AGN 
accretion event or a TDE.

The optical light curves of TDEs can be described with a model of a 
Gaussian rise plus a Power-Law (PL) decline (e.g., \citealt{van+21}),
\begin{equation}\label{tde}
F(t)= F_{p}\left\{ \begin{array}{ll}
e ^{-(t-t_{p})^2 / 2\sigma_t^2}  &  \mbox{$t \leq t_{p}$}\ \ , \\
(\frac{t-t_{p}+t_0}{t_0})^\alpha  &  \mbox{$t > t_{p}$\ \ ,}
\end{array}
\right.
\end{equation}
where flux density $F(t)$ reaches the peak $F_p$ at time $t_p$ 
and $t_0$ serves as a normalization parameter.
Using this model, we employed the Markov Chain Monte Carlo (MCMC) method, 
implemented through a python package {\tt emcee} \citep{emcee}, to fit 
the two optical flares. Because of the lack of the data points around 
the flare peaks,
we fixed $t_{p}$ of the two flares at the values estimated from the mock 
light curves (see Section \ref{mosfit}; the peak times are
MJD 57187.4 for flare~1 and MJD~59809 for flare~2, respectively). 
For flare~1, 
the Gaia Archive data were used and both rise and decline phases were fitted.
For flare~2, we used the ZTF data and only fitted the decline phase 
(i.e., $t > t_{p}$) since there were no data points on the rise phase. 
The main fitting results (90\% uncertainties) are given in 
Table~\ref{tab:lcf} and the model fits to the light curves are shown in
Figure~\ref{fig:lc_mcmc}. 
For flare~1 and flare 2 in $zg$-band, the $\alpha$ values obtained are 
consistent with the typical value of $-$5/3 within the uncertainties.
However, for flare 2 in $zr$-band, the value is slightly different from 
$-$5/3 (given the relatively large uncertainties).

\begin{table}
    \centering
    \caption{Results from light-curve fitting}
    \begin{tabular}{lccc}
        \hline
        \hline
        Band & $F_p$ & $t_p$ & $\alpha$\\
        ~ & (mJy) & (MJD) & \\
        \hline
        ~ & &  Flare 1 & \\
        \hline
        $Gaia~G$ & 0.13$^{+0.02}_{-0.01}$ & 57187.4$^*$ & $-$1.32$^{+0.32}_{-0.26}$ \\
        $W1$ & 0.89$^{+0.04}_{-0.03}$ & 57415$^{+25}_{-25}$ & $-$1.18$^{+0.13}_{-0.09}$\\\hline
        ~ & &  Flare 2 & \\
        \hline
        $zg$ & 0.09$^{+0.01}_{-0.01}$ & 59809$^*$ & $-$1.87$^{+0.63}_{-1.1}$\\
        $zr$ & 0.14$^{+0.01}_{-0.01}$ & 59809$^*$ & $-$0.84$^{+0.16}_{-0.25}$ \\
        $W1$ & 1.15$^{+0.03}_{-0.03}$ & 59981$^{+12}_{-12}$ & $-$1.15$^{+0.74}_{-0.52}$\\
        \hline
    \end{tabular}
    \tablecomments{$^*$ marks the parameters fixed in the fitting.}
    \label{tab:lcf}
\end{table}

\begin{figure*}
    \centering
    \includegraphics[width=0.48\linewidth]{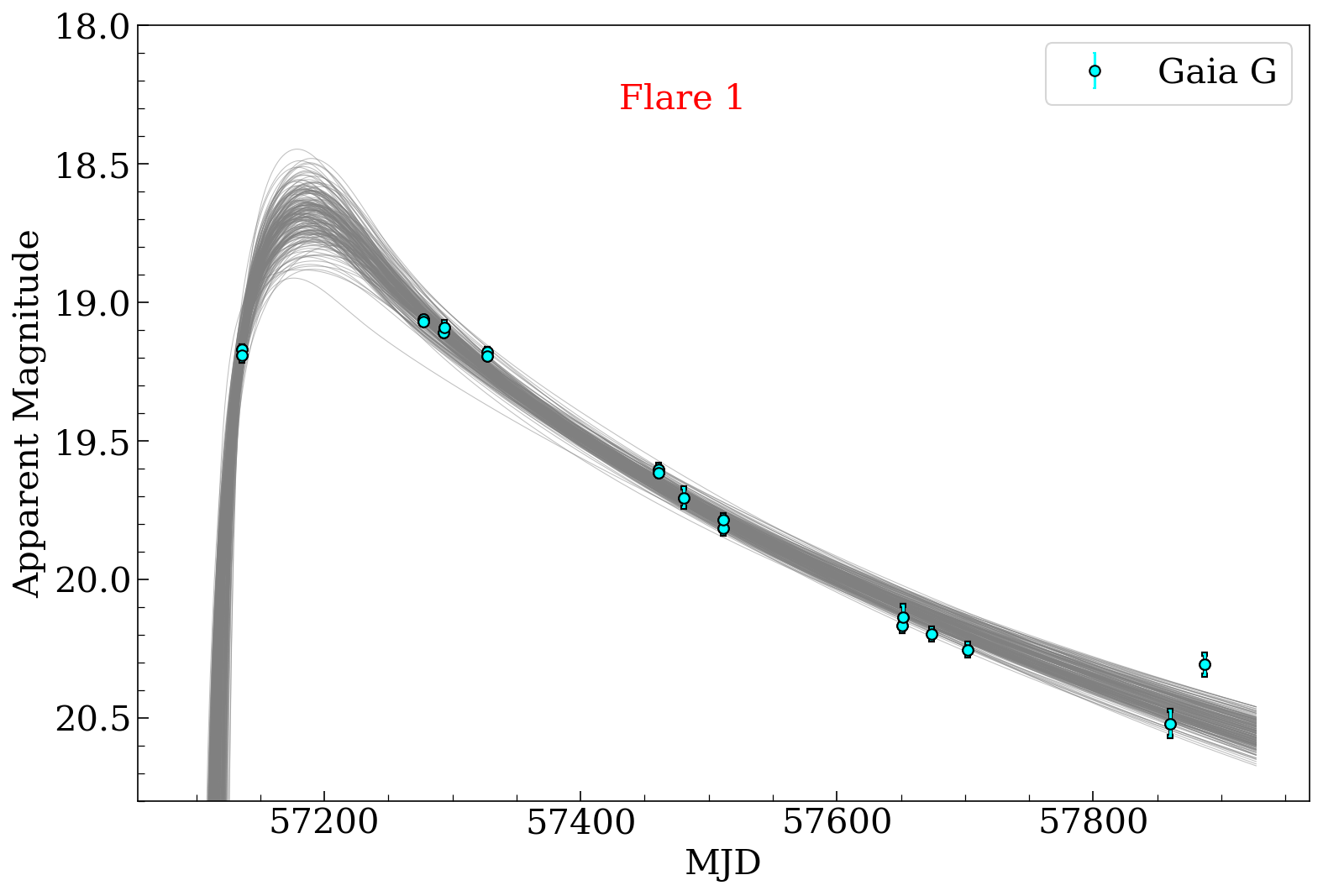}
    \includegraphics[width=0.50\linewidth]{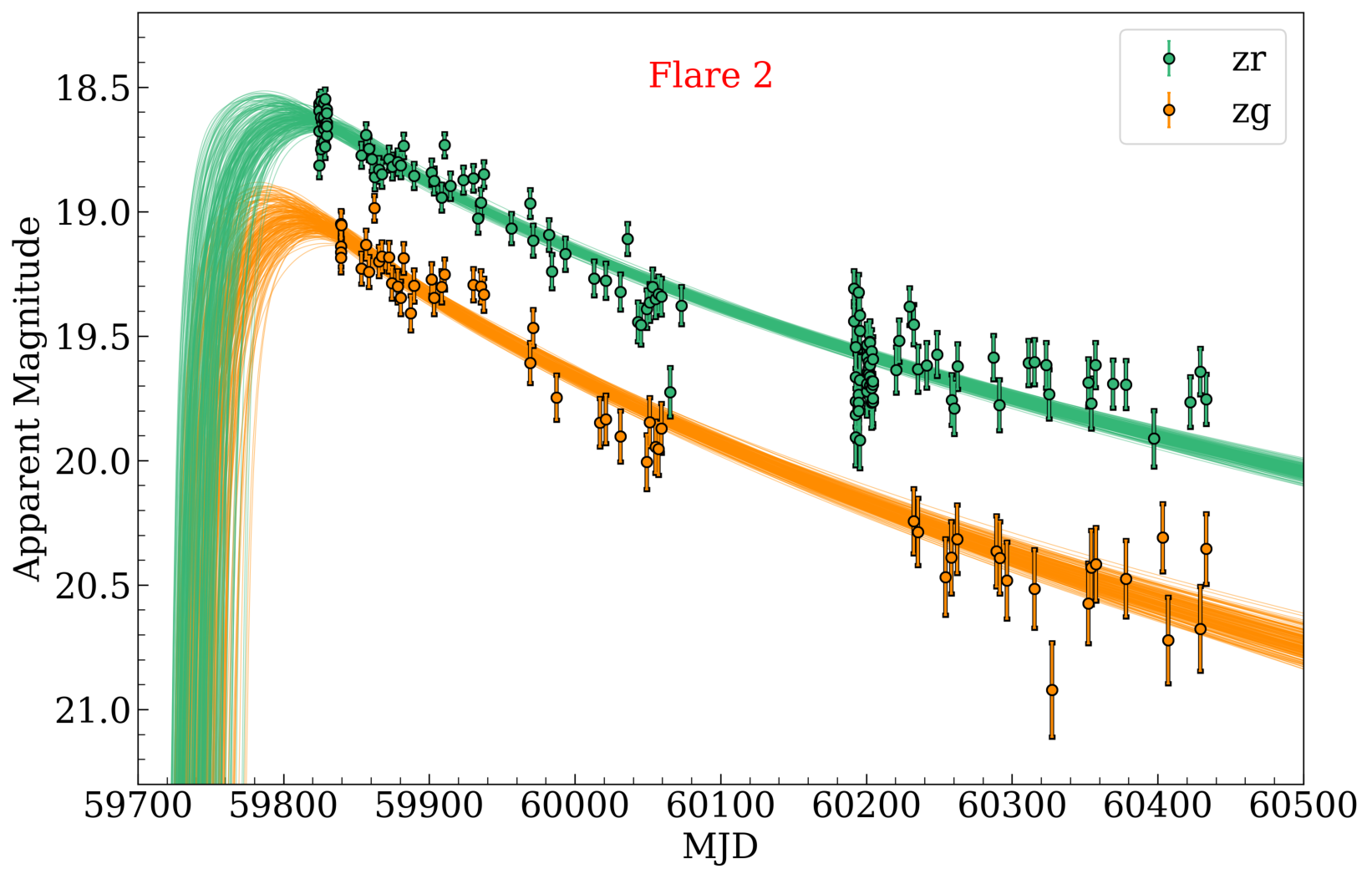}
	\caption{Mock light curves for flare~1 ({\it left}) and flare~2 
	({\it right}) of AT2022sxl, generated by {\tt MOSFiT} using the
	TDE model. }
    \label{fig:mosfit_lc}
\end{figure*}

We also performed the fitting to two MIR flares with the model 
(cf., Eq.~\ref{tde}) to estimate their peak times.
Here, we set $t_{p}$, the peak time at the MIR bands,
as a free parameter and used the W1 data. The obtained results are also
given in Table~\ref{tab:lcf}.
Comparing the results, we found that 
the MIR peak flux of flare~2 is slightly higher than that
of flare~1, and the time delays of the MIR peaks with respect to the
respective optical 
peaks are $\sim$228\,d for flare~1 and $\sim$172\,d for flare~2. 
Considering the uncertainties of both optical and MIR peak times, the two 
delays are compatible.

\subsubsection{TDE model fitting in {\tt MOSFiT}}\label{mosfit}

Assuming AT2022sxl is a rTDE, we also employed a Python-based package, 
the Modular Open Source Fitter for Transients 
({\tt MOSFiT}, \citealt{Guillochon+18}), to fit the two flares with
the TDE model, which was described in detail in \citet{Mockler+19}. 
The fitting allowed us to probe physical parameters and obtain
their possible value ranges.
The model 
has 10 free parameters (Table~\ref{tab:mf}), which are the masses of 
the black hole $M_{\rm BH}$ and the disrupted star $M_{\ast}$, 
the scaled impact parameter $b$ 
($\geq$ 1 or $<1$ respectively implies full or partial disruption), 
the efficiency $\epsilon$ of converting fallback rate to luminosity, 
the viscous delay time $t_{\rm vis}$, 
the PL exponent $l_{ph}$ and the radius normalization $R_{ph0}$ of 
the photosphere, the time of first fallback, $t_{exp}$, since the first 
detection, the hydrogen column density $n_{\rm H}$ in the host galaxy, 
and a white noise parameter $\sigma_n$. We fixed the luminosity distance
at 1.18\,Gpc (from $z = 0.23$) and the Galactic reddening $E(B-V) = 0.043 $ 
(from \citealt{sf11}). The priors of $t_{exp}$ could be 
limited by the last observations before the two flares ($-$40 d for flare~1 
and $-$110\,d for flare~2). The priors of all parameters are listed in 
Table~\ref{tab:mf}. We used Gaia Archive $G$-band data for fitting 
flare~1 and $zg$- and $zr$-band data for fitting flare~2. 
We performed the fitting until convergence using the {\tt DYNESTY} 
package, which uses the dynamic nested sampling \citep{Speagle+20}. 
The posterior distributions of the model parameters from the fitting
are show in Appendix Figure~\ref{fig:corner} and the 
obtained parameter values are given in Table~\ref{tab:mf}.
The mock light curves are shown in Figure~\ref{fig:mosfit_lc}.

\begin{table}
    \centering
    \caption{MOSFiT TDE fitting results}
	\label{tab:mf}
    \begin{tabular}{lccc}\hline
Parameter & Prior & Flare~1 & Flare~2 \\
        \hline
        log($M_{\rm BH}$)/$M_\odot$ &  [6.7, 8.7] & 7.16$^{+0.09}_{-0.14}$ & 7.07$^{+0.13}_{-0.15}$ \\
        $M_{\ast}$/$M_\odot$ & [0.01, 100] & 0.38$^{+0.23}_{-0.15}$ & 0.35$^{+0.16}_{-0.10}$ \\
        $b$ & [0, 2] & 0.96$^{+0.17}_{-0.12}$ & 1.00$^{+0.04}_{-0.05}$ \\
        $\log(\epsilon)$ & [$-$2.3, $-$0.4] & $-$0.83$^{+0.21}_{-0.24}$ & $-$0.67$^{+0.13}_{-0.14}$ \\
        log($R_{ph0}$) & [$-$4, 4] & 0.33$^{+0.16}_{-0.14}$ & 0.57$^{+0.11}_{-0.09}$ \\
        $l_{ph}$ & [0, 4] & 0.27$^{+0.07}_{-0.10}$ & 0.09$^{+0.05}_{-0.04}$ \\
        log($t_{\rm vis}$)/d & [$-$3, 3] & $-$2.01$^{+0.94}_{-0.63}$ & $-$1.2$^{+1.3}_{-1.3}$ \\
        $t_{exp}$/d & [$-$40/$-$110, 0] & $-$33.6$^{+6.4}_{-4.4}$ & $-$95$^{+13}_{-10}$ \\
	    log($n_{\rm H})$/cm$^{-2}$ & [19, 23] & 21.02$^{+0.22}_{-0.30}$ & 21.35$^{+0.04}_{-0.05}$ \\
        log($\sigma_n$) & [$-$4, 2] & $-$1.08$^{+0.09}_{-0.08}$ & $-$1.11$^{+0.05}_{-0.04}$ \\
        \hline
    \end{tabular}
\tablecomments{Uncertainties of the parameters are the 16th and 84th percentiles (1$\sigma$)}
\end{table}

From the fitting to the flare~1 data, we found that a star of 
$M_{\ast} \approx 0.38$\,$M_{\odot}$ was disrupted by an SMBH of 
$M_{\rm BH} \approx 10^{7.16}$\,$M_{\odot}$ with the peak time at
MJD~57187.4$\pm$5.0. The scaled impact parameter $b \approx 0.96$, which is
slightly smaller than 1 and may suggest a partial disruption (but note that
the uncertainty is large).
For flare 2, similar parameter values were obtained.
It was an event of a star of $M_{\ast} \approx 0.35$\,$M_{\odot}$ being 
disrupted by an SMBH of $M_{\rm BH} \approx 10^{7.07}$\,$M_{\odot}$, with the 
peak time at MJD~59809$^{+12}_{-13}$. The scaled impact parameter 
$b \simeq 1.00$, which corresponds to a full disruption. The low mass
of the disrupted star and $b\sim 1$ in flare 1 do not support a rTDE case. 
However, it should be noted that the fitting contains large systemic 
uncertainties on parameters (e.g., $\pm0.35$ for $b$ and $\pm0.66M_{\odot}$ 
for $M_{\ast}$; \citealt{Mockler+19}); whether a repeating case or not
requires spectroscopic identification. 
The peak times
indicate a separation time of $\sim$7.2\,yr between the two flares, 
within the range of the currently reported (candidate) rTDEs.

\begin{figure*}
	\centering
	\includegraphics[width=0.79\linewidth]{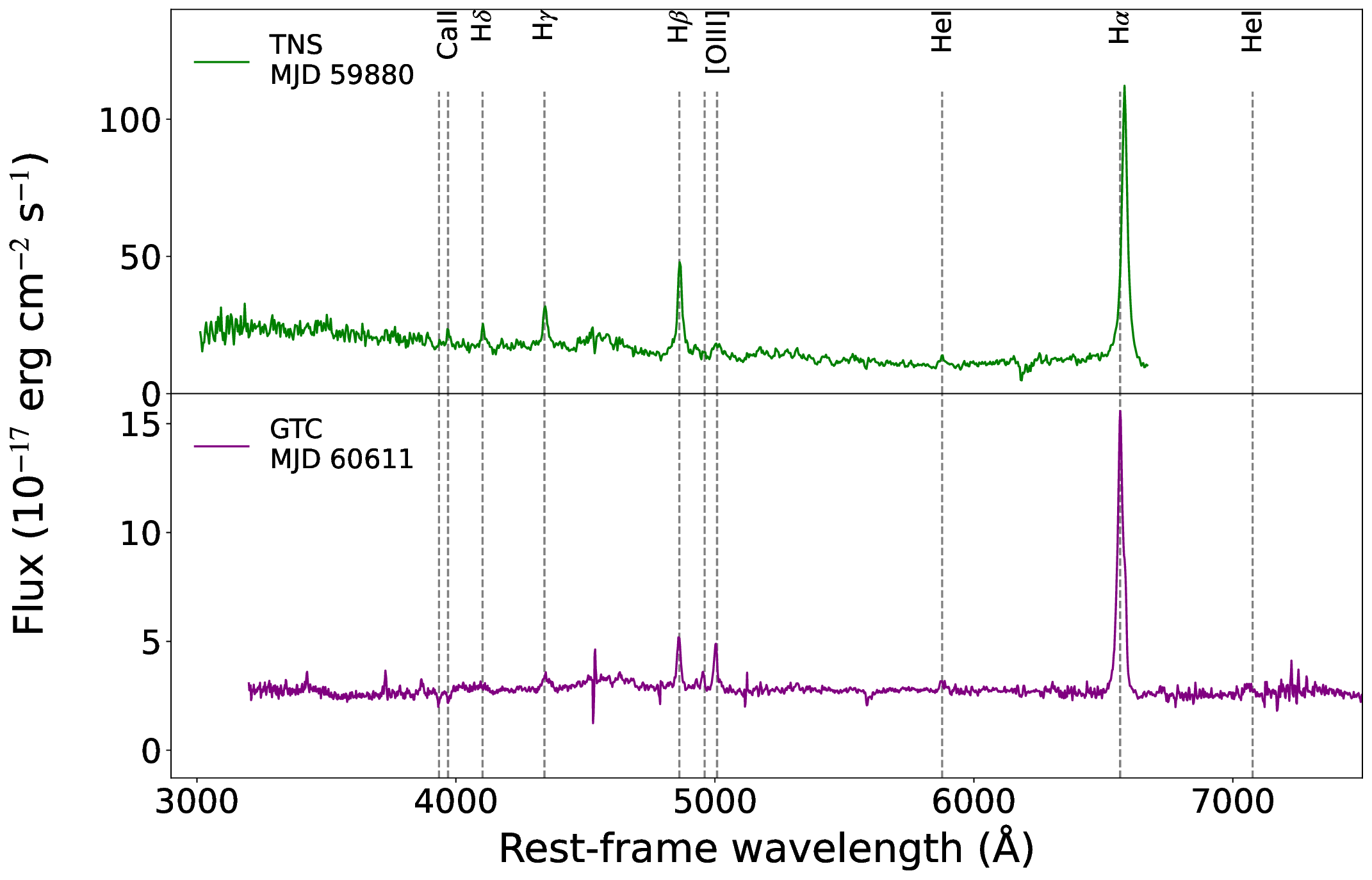}
	\caption{TNS ({\it upper}) and GTC ({\it lower}) spectra of 
	AT2022SXL. Prominent emission and absorption features are marked.}
	\label{fig:spec}
\end{figure*}

\subsection{Optical Spectrum Data}

\subsubsection{Data and observation}

There is one spectrum, taken on Oct. 28, 2022 (MJD~59880),
of the
transient AT2022sxl reported at the Transient Name Server (TNS)\footnote{\url{https://www.wis-tns.org/}}. 
The observation time is approximately 71\,d after the optical peak of 
flare~2 (Figure~\ref{fig:multi}). This TNS spectrum shows a strong continuum 
and prominent emission lines in the H$\alpha$ and H$\beta$ regions 
(Figure~\ref{fig:spec}). The redshift $z$ of the source was determined
to be $z = 0.23$. 

We also obtained a spectrum of the source on Oct. 27, 2024 (MJD~60611)
with the 10.4-m Gran Telescopio Canarias (GTC),
at the Observatorio del Roque de los Muchachos in La Palma, Spain, 
when the source was back to its quiescent flux level according to 
the light curves (Figure~\ref{fig:multi}). The observation was conducted 
with the Optical System for Imaging and low-intermediate Resolution 
Integrated Spectroscopy (OSIRIS; \citealt{osiris00}). This instrument is 
equipped with 
one $4096 \times 4096$ pixel$^2$ Marconi CCD. A standard 
$2 \times 2$\,pixel$^2$ binning was set in our observation, which provided 
a plate 
scale of $0\farcs254$\,pixel$^{-1}$. We used a long slit of width 
1\farcs0. The grisms were R1000R and R1000B, and the first (second) provided
a wavelength coverage of approximately 5100\,\AA--10000\,\AA\ 
(3630\,\AA--7500\,\AA) with a spectral dispersion 
of 2.62\,\AA\ pixel$^{-1}$ (2.12\,\AA\,pixel$^{-1}$). 
Three exposures of 400\,s each were taken with R1000R and two exposures of 
1000\,s each were taken with R1000B.

The raw imaging data were reduced using the Image Reduction and Analysis 
Facility (IRAF)\footnote{\url{https://iraf-community.github.io/}} packages.
The reduction included bias subtraction and flat-fielding correction.
The multiple exposures of one same grism setting were combined.
Wavelength calibration was performed using spectra of a HgAr comparison lamp
taken during the observation run. Flux calibration was achieved using 
observations of the spectrophotometric standard star Feige 110.

We further combined the two spectra into one by averaging their overlapping
regions. Since the spectral dispersions are different, linear interpolation
of the region of the blue spectrum on the wavelength grids of the red spectrum 
was performed. The final spectrum is shown in
Figure~\ref{fig:spec}. Nearly the same line features as those in the TNS
spectrum were detected, but with weaker strengths.

\subsubsection{Spectral analysis}\label{spec}

We analyzed the TNS and GTC spectra using the PyQSOFit software 
package \citep{gsw18}, a Python-based tool designed for fitting spectra 
of quasars. In the fitting, we found that the continuum of 
a host galaxy was needed. This is supported by the appearance of 
the Ca II doublet (i.e., $\lambda$3934 and $\lambda$3969) absorption
feature in
the GTC spectrum. The parameters of the prominent emission lines
derived from the fitting are summarized in Table~\ref{tab:lines}. 
Details of the fitting for each spectrum are shown
in Appendix Figure~\ref{fig:sf1}. 

\begin{table}
	\centering
	\caption{Measurements of emission features in the two spectra}
	\label{tab:lines}
	\begin{tabular}{lcc}
		\hline
		Line & TNS & GTC \\
		\hline
		
		\multicolumn{3}{l}{H$\beta$\quad broad} \\
		\hline
		FWHM & $2825 \pm 35$ & $ 2270 \pm 200$ \\
		EW       & $53.9 \pm 9.5$ & $28.3 \pm 2.6$ \\
		Flux  & $620 \pm 110$ & $35.2 \pm 3.3$ \\
		\hline
		
		\multicolumn{3}{l}{H$\beta$\quad narrow} \\
		\hline
		FWHM  & $973 \pm 48$ & $946 \pm 25$ \\
		EW       & $49.3 \pm 5.7$ & $38.5 \pm 2.2$ \\
		Flux  & $569 \pm 66$ & $47.8 \pm 2.8$ \\
		\hline
		
		\multicolumn{3}{l}{[OIII] $\lambda$4959\quad narrow} \\
		\hline
		FWHM  & --- & $945 \pm 25$ \\
		EW       & --- & $11.08 \pm 0.46$ \\
		Flux & --- & $13.45 \pm 0.56$ \\
		\hline
		
		\multicolumn{3}{l}{[OIII] $\lambda$5007\quad narrow} \\
		\hline
		FWHM  & --- & $945 \pm 25$ \\
		EW      & --- & $35.9 \pm 1.8$ \\
		Flux  & --- & $43.1\pm 1.0$ \\
		\hline
		
		\multicolumn{3}{l}{He I\quad broad} \\
		\hline
		FWHM & $2830 \pm 850$ & $2828.6 \pm 1.3$ \\
		EW       & $18 \pm 10$ & $11.9 \pm 3.3$ \\
		Flux  & $77 \pm 43$ & $12.6 \pm 3.5$ \\
		\hline
		
		\multicolumn{3}{l}{He I\quad narrow} \\
		\hline
		FWHM  & $1190 \pm 200$ & $1200 \pm 260$ \\
		EW       & $22.6 \pm 6.3$ & $11.5 \pm 3.5$ \\
		Flux  & $96 \pm 26$ & $12.3 \pm 3.7$ \\
		\hline
		
		\multicolumn{3}{l}{H$\alpha$\quad broad} \\
		\hline
		FWHM  & $2830 \pm 170$ & $2820 \pm 210 $ \\
		EW       & $460 \pm 82$ & $99.3 \pm 7.7$ \\
		Flux  & $2100 \pm 380$ & $92.4 \pm 7.2$ \\
		\hline
		
		\multicolumn{3}{l}{H$\alpha$\quad narrow} \\
		\hline
		FWHM  & $962 \pm 21$ & $776.4 \pm 8.2$ \\
		EW       & $464 \pm 45$ & $386.1 \pm 3.4$ \\
		Flux  & $2130 \pm 210$ & $265.4 \pm 3.1$ \\
		\hline

		[Ne II] $\lambda$6585 flux & --- & $115.9  \pm 2.4$ \\
		\hline
		
		Fe~II flux & $1590 \pm 170$ & $86.0 \pm 6.4$ \\
		\hline
		log($L_{5100}$) (erg s$^{-1}$) & $43.871 \pm 0.018$ & $42.974 \pm 0.056$ \\
		\hline
		log($M_{BH}$/$M_{\odot}$) & $7.75^{+0.19}_{-0.24}$ & $7.43^{+0.12}_{-0.13}$ \\
		\hline
	\end{tabular}
	
	\tablecomments{1) Full Width at Half Maximum (FWHM), Equivalent Width (EW),
	and Flux (including Fe~II) are in units of km s$^{-1}$, \AA, 
	and $10^{-17}$ erg cm$^{-2}$ s$^{-1}$, respectively. 
	2) Approximate systematic uncertainties of 19\% and 10\% for the TNS 
	and GTC spectrum, respectively,
	should be considered for the measurements of the line features. 3) The 
	systematic uncertainties are included to estimate the black hole masses.}
\end{table}

By comparing the results from fitting the two spectra, we found broad
components in the lines of H$\alpha$, He I $\lambda$5876, and H$\beta$. They
nearly all had Full Width at Half Maximum (FWHM) values 
of $\sim$2800\,km\,s$^{-1}$ (except H$\beta$ in the GTC spectrum),
but the Equivalent Widths (EWs) decreased significantly from 
the flaring to quiescent state.
The corresponding narrow components
mostly had FWHMs of $\sim$900\,km\,s$^{-1}$, and the EWs also decreased
in the quiescent state
but not as significantly as the broad components. It is likely that 
the putative TDE mainly caused the changes of the broad components. 
Using the fitting results from the GTC spectrum, we found that
the flux ratios of log[NII]/H$\alpha \simeq -0.36$ and 
log[OIII]/H$\beta \simeq 0.05$. The ratios put this host at the region
of the composite galaxies in the BPT diagram 
(e.g., \citealt{kht+03, knt+17}). Therefore, the weak broad
components in the GTC spectrum could be the residual emission of the TDE event
(in other words, they did not totall disappear although the continumm was 
back to the quiescent flux level).
We also detected the Fe II emission (note that because of it,
we could not obtain proper measurements of the [O III] lines in
the TNS spectrum), and similarly its 
flux in the GTC spectrum decreased significantly.
In addition, as the He I $\lambda$5876 line changed accordingly,
it suggests that this is an H+He TDE \citep{gez21}.

We estimated the central black hole mass ($M_{\mathrm{BH}}$) using the 
measurements of the TNS spectrum.  Specifically, given the
FWHM of the broad component of H$\beta$ FWHM(H$\beta$) and
the continuum luminosity at the rest-frame wavelength of 5100\,\AA\,
$L_{5100}$, following \citet{vp06},
\begin{equation}
	\log \left( \frac{M_{\mathrm{BH}}}{M_{\odot}} \right) = \log \left[ \left( \frac{\mathrm{FWHM}(\mathrm{H}\beta)}{\mathrm{km\,s}^{-1}} \right)^{2} \left( \frac{\lambda L_{5100}}{10^{44}\,\mathrm{erg\,s}^{-1}} \right)^{0.5} \right] + 0.91\ \ .
	\label{eq:mbh_hb} 
\end{equation}
$M_{\mathrm{BH}} \simeq 10^{7.75}$\,$M_{\odot}$
(or even based on the measurements of the GTC spectrum,
$M_{\mathrm{BH}} \simeq 10^{7.43}$\,$M_{\odot}$). The detailed values are 
given in Table~\ref{tab:lines}, for which we included the systematic 
uncertainties of 19\% and 10\% for the TNS and GTC spectrum respectively.
The uncertainties were estimated by selecting multiple continuum regions 
of similar flux levels, calculating
the average fluxes of these regions and comparing them to the average flux 
of the region with the lowest noise level, and obtaining the mean difference 
(the systematic uncertainty) between these averages for each spectrum. 

\begin{figure}
    \centering
    \includegraphics[width=0.99\linewidth]{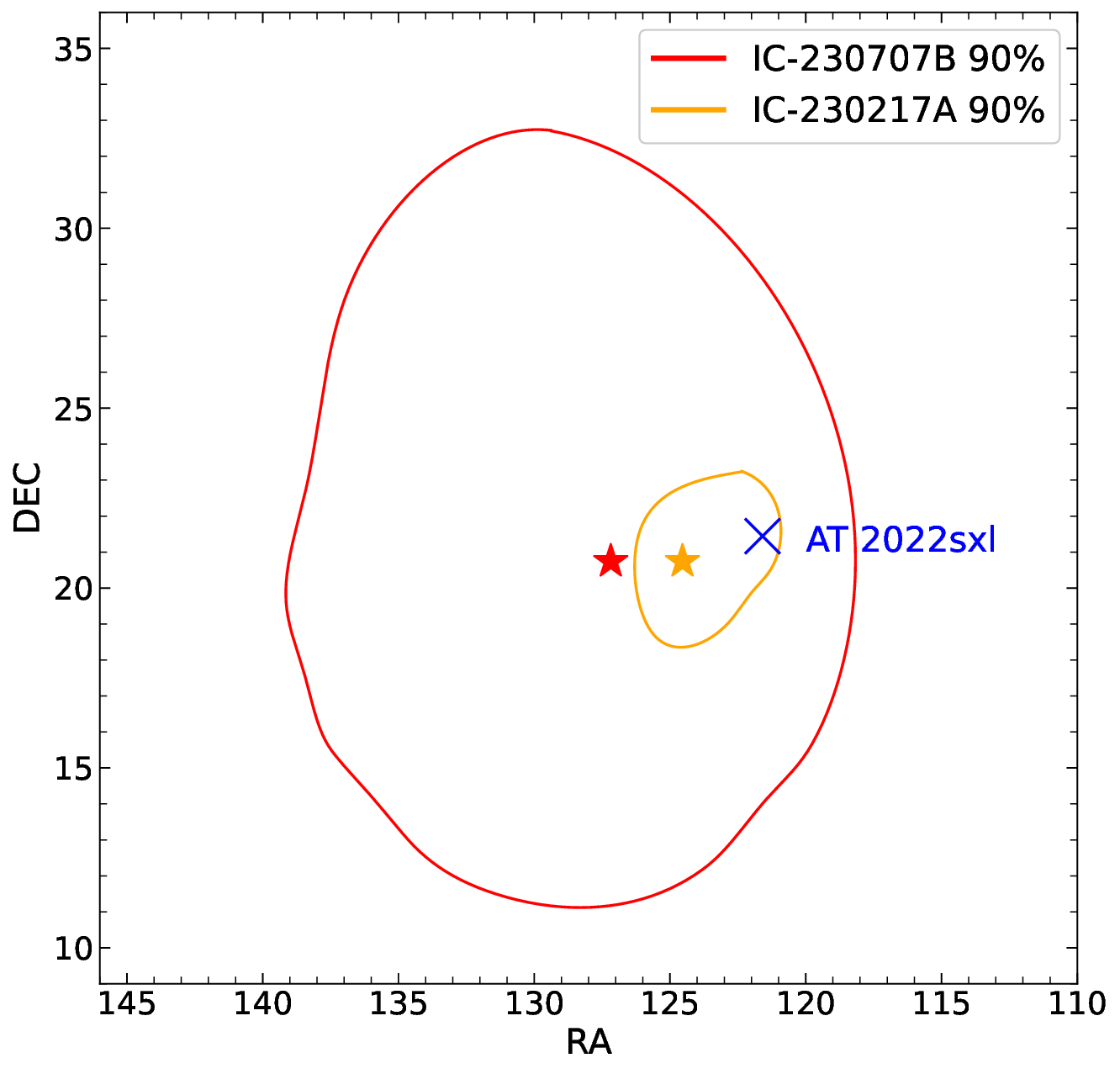}
	\caption{Positional uncertainty (at a 90\% confidence level) regions 
	of IC-230217A (orange) and 
	IC-230707B (red) given in IceCat-1. The blue cross marks the
	position of AT2022sxl.}
    \label{fig:neu}
\end{figure}
	
\subsection{Positional and Temporal Coincidences with Two High-energy Neutrinos}\label{pos}
We noted that two HE neutrino events detected by IceCube, 
IC-230217A \citep{IC230217A} and IC-230707B \citep{IC230707B}, were in 
positional coincidence with AT2022sxl 
(see Figure \ref{fig:neu}). Furthermore, they arrived during flare 2, in
particular around and after the peak of the MIR variations
(see Figure \ref{fig:lc_mcmc}). These two neutrinos were also listed in 
IceCat-1 \citep{Abbasi+23}. IC-230217A was a Bronze type event with an energy 
of $\sim$55\,TeV and a signalness of $\sim$45.4\% (the probability of having 
an astrophysical origin). It was detected on 2023 February 17 20:49:43.376 UT 
(MJD~59992.87) with a position of R.A. = 124\fdg54$^{+1\fdg58}_{-3\fdg52}$, 
Decl. = 20\fdg74$^{+2\fdg38}_{-2\fdg49}$ (equinox J2000.0, 90\% uncertainty).
The second one, IC-230707B, also a Bronze type event with an energy 
of $\sim$154\,TeV and a signalness of $\sim$ 46.6\%, was detected on 2023 
July 7 18:56:51.435 UT (MJD~60132.79). Its position, having large uncertainties,
was R.A. = 127\fdg18$^{+10\fdg63}_{-8\fdg96}$, 
Decl. = 20\fdg74$^{+9\fdg25}_{-9\fdg95}$ (equinox J2000.0, 90\% uncertainty).
The time delays of the arrival times of the two neutrinos with respect to 
the optical peak time of flare 2 are $\sim$184\,d and $\sim$324\,d, 
corresponding
to 150\,d and 263\,d in the SMBH frame, respectively. The time delays 
with respect to MIR peak time are $\sim$ 12\,d and $\sim$152\,d,
corresponding to 10\,d and 123\,d in the SMBH frame, respectively. 
This better temporal match with the MIR peak than with the optical peak
is a common feature of previously reported neutrino TDEs. In addition, 
the neutrinos in the previous TDE cases had time delays of 100--300\,d 
relative to the optical peaks (e.g., \citealt{van+24}), and
the AT2022sxl neutrino case is in the range.

To quantify the association, we estimated the probability by 
coincidence between AT2022sxl and the two neutrinos. In 
IceCat-1 \citep{Abbasi+23}, the rate of Bronze-type events per unit time
can be estimated as 
$n_s =(N_s/T_{\rm IceCube})\times(\Omega_{n}/4\pi)\approx 0.02$\,yr$^{-1}$, 
where $N_s=222$ is the number of Bronze-type events, 
$T_{\rm IceCube}\simeq 12$\,yr is 
the live time of IceCube and $\Omega_{n}$ is the average containment area of 
a Bronze-type event. $\Omega_{n} \simeq \Omega_{sum}/N_s$, where 
$\Omega_{sum}$ is the 90\% containment area of all Bronze-type events (for 
simplicity, the containment region of each event is approximated as an ellipse).
The expected neutrino number of coincidences is obtained by multiplying $n_s$ 
by the time interval of $\sim7.6$\,yr of the MIR flares 
(i.e., MJD 57415--60200). The obtained value is $\simeq$0.13 and the Poisson 
probability to see at least two Bronze type events is $\simeq$0.8\%.

\section{Discussion}

From collecting and analyzing archival data, we have found two
continuous flares, with a separation time of $\sim$7.2\,yr, in a
$z = 0.23$ composite galaxy (Figure~\ref{fig:lc_mcmc}). The flares appeared 
to have
similar peak fluxes and $\sim$600\,d lasting times and were accompanied with 
a delayed MIR flare.  
We only obtained two spectra of the source,
one near the optical peak of flare 2 and the other at the quiescent flux level
after flare 2. 
The two spectra showed prominent emission features. Broad components
in the lines of H$\alpha$, H$\beta$, and He~I $\lambda$5876 were detected,
with most of them having FWHMs $\sim$2800\,km\,s$^{-1}$.
The significant EW decreases
of the broad components are seen from the TNS to GTC spectrum,
suggesting that the flare was the cause of the broad components. 
The much weaker broad components in the GTC spectrum thus likely were
the residual emission of the flare, although the continuum was already
back to the quiescent flux level.

On the basis of these and the flares' shape, having a fast rise and 
an $\alpha\sim -5/3$ PL decay, we are inclined to identify AT2022sxl as a candidate rTDE. However,
the fitting with the TDE model in MOSFiT did not provide ideal results for
a rTDE. Also, no spectra are available for flare 1,
and thus no verification for the repeating nature could be made.
Alternatively, binary SMBH systems are thought
to be able to give rise to periodic-like flux variations 
(e.g., \citealt{pay+21} and references therein). However for
such a binary, sinusoidal-like
periodic modulation is often considered for the flux variations
(e.g., \citealt{gra+15},
\citealt{li+23} and references therein). In any case, if this is a potential
binary SMBH system, a third flare may be expected to appear in near future. 
The other possibility
for the flares is activities related to AGN accretion, or generic AGN flares 
(e.g., \citealt{gra+17,tra+19}). Given the GTC spectrum, AT2022sxl would be 
an AGN turn-on case.  Such cases are
rare, and the characteristics of
related flux variations are not clear and the variation timescales could be 
long according to the case reported in \citet{are+24}. We checked X-ray and
radio detections at the source position for more information, but no
X-ray observations were available during the two flares and no radio sources
were reported at the position. Below, we discuss AT2022sxl
within the TDE scenario for its association with the neutrinos. We note that
because of the detection of He~I lines, this candidate would be
an H+He TDE.


\begin{table*}

    \centering

      \caption{Luminosity properties of the (candidate) neutrino-emitting TDEs}

      \label{tab:tdes}

\begin{tabular}{lcccccc}

        \hline

        Source & Neutrino & $z$ & log$L_{OUV}^{peak}$ & log$L_{W1}^{peak}$ & Reference \\

        ~ & ~ & ~ & ($\rm erg~s^{-1}$) & ($\rm erg~s^{-1}$) & \\

        \hline

      AT2019dsg & IC-191001A & 0.051 & 44.4 & 43.0 & 1, 2, 8 \\

        AT2019fdr & IC-200530A & 0.267 & 45.1 & 44.3 & 3, 8 \\

        AT2019aalc & IC-191119A & 0.036 & 44.4 & 43.5 & 2, 8 \\

        SDSSJ104832.79+122857.2 & IC-200109A & 0.0537 & --- & 42.8 & 4 \\
        SDSSJ164938.77+262515.3 & IC-200530A & 0.0588 & --- & 43.1 & 4 \\
        AT2021lwx & IC-220405B & 0.995 & 46.1 & 44.9 & 5, 6 \\

        ATLAS17jrp & Neutrino flare & 0.0655 & 44.2 & 43.3 & 7 \\

        AT2022sxl  & IC-230217A, IC-230707B & 0.23 & 45.0 & 44.2 & 9 \\

        \hline

    \end{tabular}

      \tablecomments{The peak bolometric luminosity $L_{OUV}^{peak}$ of AT2022sxl is obtained from {\tt MOSFiT}; The references are 1. \citet{ste+21}; 2. \citet{van+24}; 3. \citet{reu+22}; 4. \citet{jia+23}; 5. \citet{wis+23}; 6. \citet{ywl24}; 7. \citet{li+24}; 8. \citet{Winter+23}; 9. this work.}

\end{table*}

In this putative TDE, delayed MIR flares were observed. This so-called 
`dust echo' is understood as the strong optical--to--X-ray emission powered by
a TDE is re-processed by surrounding dust, resulting in a delayed MIR flare
or enhanced MIR emission for an AGN host (see, e.g., \citealt{jia+21, van+21}).
The delayed time $\delta t$ in AT2022sxl was $\sim$228\,d for flare 1 
and $\sim$172\,d for flare 2 (note that the uncertainties on $\delta t$
are large, making the two $\delta t$ values compatible with each other).
The values are in the range of those detected in previously reported TDEs
(e.g., \citealt{jia+21}). For reported TDEs in possible association with
neutrinos, one feature is the arrival times of neutrinos match better with
the peak times of MIR flares than with optical peaks (the latter are considered
to reflect the peak mass accretion rate; e.g., \citealt{Mockler+19}). 
Taking AT2022sxl as an example, the first neutrino time
is nearly at the MIR peak (considering the uncertainty). 
Therefore, it is possible that the neutrino production is closely related
to MIR flares (see \citealt{ywl24} and references therein). HE protons
are accelerated in a TDE event and interact with MIR photons (i.e.,
the $p\gamma$ process), and thus the peak time of the latter determines
the time delay of neutrino events with respect to the peak time of
this TDE \citep{yw23}. 

\begin{figure}
    \centering
    \includegraphics[width=0.99\linewidth]{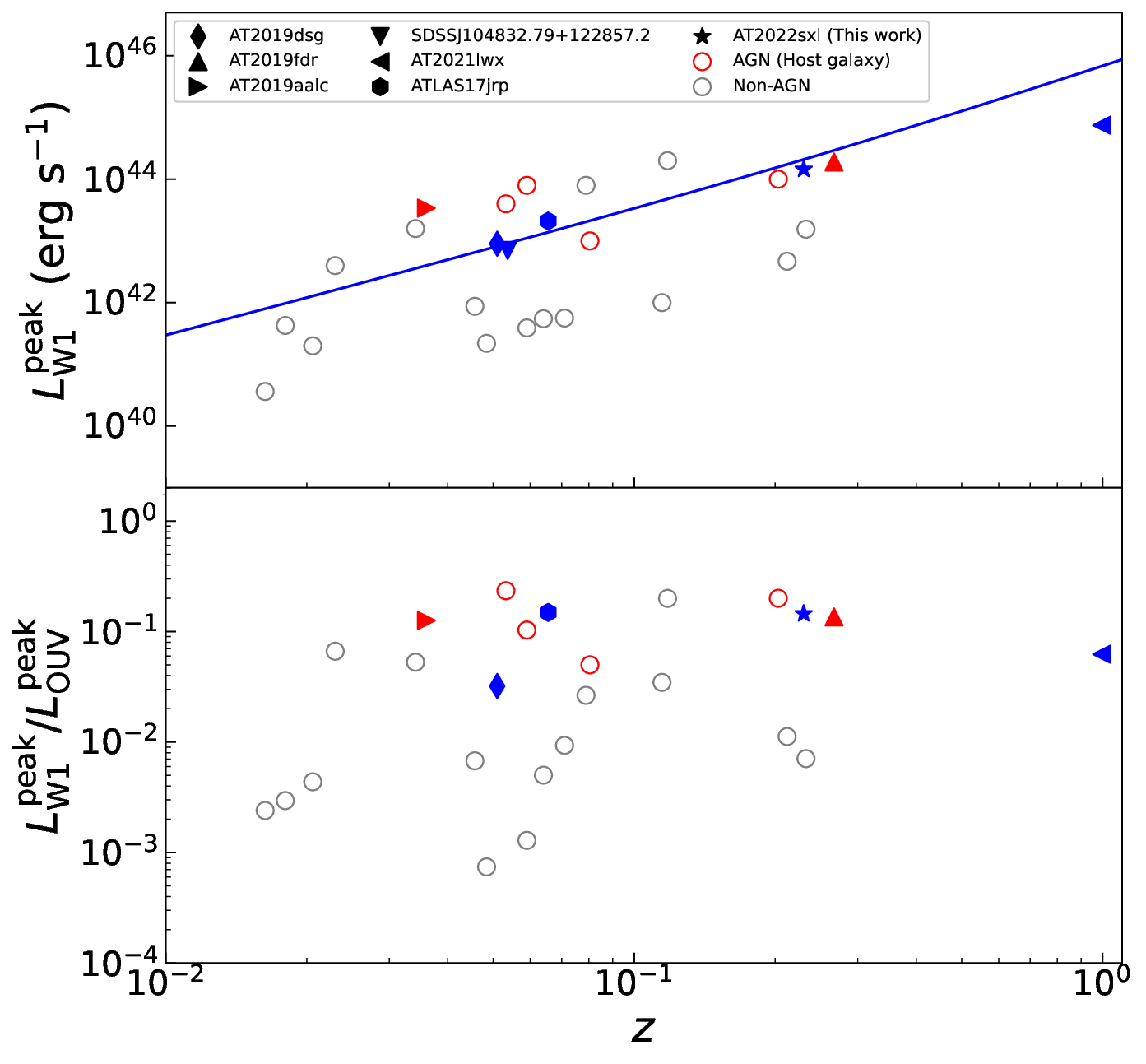}
	\caption{{\it Top:} MIR W1 peak luminosities 
	of the reported possible neutrino-emitting TDEs. Also shown for
	comparison are MIR peak luminosities (circles) of TDEs listed 
	in \citet{jia+21} and \citet{van+21b}. TDEs in an AGN-host are marked 
	with red, which include two neutrino ones. 
	For most of the neutrino-emitting TDEs, an average W1 peak flux of
	1.2$\times 10^{-12}$\,erg\,s$^{-1}$\,cm$^{-2}$ is obtained 
	(blue line). {\it Bottom:} peak W1-to-OUV luminosity ratios of 
	the TDEs in the top panel (note that SDSS J104832.79+122857.2 did not 
	have an optical counterpart and so is not shown).}
    \label{fig:mirl}
\end{figure}

In Table~\ref{tab:tdes}, we collected the MIR W1 luminosities 
$L^{\rm peak}_{\rm W1}$ of the
reported eight neutrino-emitting TDEs for comparison.
These TDEs' $L^{\rm peak}_{\rm W1}$ versus $z$, including
AT2022sxl, are plotted in the top panel of Figure~\ref{fig:mirl} 
(note that SDSSJ164938.77+262515.3 and AT2019fdr have been reported 
to be associated with the same neutrino event, and we only kept the latter
in the plot). Five of them had
similar MIR peak fluxes, with an average W1 value of 
$F_{\rm W1}^a \simeq 1.2\times 10^{-12}$\,erg\,s$^{-1}$\,cm$^{-2}$ 
(the standard deviation
is $\simeq$4.5$\times 10^{-13}$\,erg\,s$^{-1}$\,cm$^{-2}$). The other two,
AT2019aalc, hosted in a Seyfert-1 galaxy, had a $\sim$8 times higher peak flux
and AT2021lwx, with the highest redshift $z=0.995$ among the TDEs, 
had a $\sim$9 times
lower peak flux. We also plotted $L^{\rm peak}_{\rm W1}$ of those TDEs 
(19 in total) reported in \citet{jia+21} and \citet{van+21b} for comparison.
As can be seen in Figure~\ref{fig:mirl}, 11 of them had significantly lower 
luminosities.
The obvious exceptions are four TDEs hosted in an AGN, which had 
comparable luminosity values, in addition to four other TDEs above 
the $F_{\rm W1}^a$ line in Figure~\ref{fig:mirl}.
The plot thus suggests the association of neutrino emission with TDEs
with more powerful MIR flares, although it should be noted that
among TDEs in Table~\ref{tab:tdes}, AT2019fdr and AT2019aalc had an AGN host
and the event reported by \citet{jia+23} was only an MIR flare. 

We further checked the peak luminosity ratios 
$L^{\rm peak}_{\rm W1}/L^{\rm peak}_{\rm OUV}$ 
(or the so-called covering factors)
for these TDEs, where
$L^{\rm peak}_{\rm OUV}$ are the optical-to-ultraviolet (OUV) luminosities
(Table~\ref{tab:tdes}). The ratios are plotted in the bottom panel of
Figure~\ref{fig:mirl} (note that the candidate neutrino TDE
SDSS J104832.79+122857.2 did not have an optical counterpart). As can be seen,
the luminous MIR TDEs in the top panel of the figure generally
have large MIR-to-OUV ratios ($\gtrsim 0.04$); in other words, 
their large MIR luminosities are not the results of having luminous OUV 
emission. The existence of sufficient dust in the vicinity of an SMBH should
play a key role for showing luminous MIR emission. It is interesting to
note that among the sources with
relatively high $L^{\rm peak}_{\rm W1}$ or high luminosity ratios, 
nearly half are 
AGNs and two of them are neutrino-associated sources. Thus, we may suspect
that luminous MIR emission is a prerequisite for neutrino production in TDEs.

\begin{acknowledgments}

This work was based on observations obtained with the Samuel Oschin Telescope 
48-inch and the 60-inch Telescope at the Palomar Observatory as part of the 
Zwicky Transient Facility project. ZTF is supported by the National Science
Foundation under Grant No. AST-2034437 and a collaboration including Caltech, 
IPAC, the Weizmann Institute for Science, the Oskar Klein Center at Stockholm 
University, the University of Maryland, Deutsches Elektronen-Synchrotron
and Humboldt University, the TANGO Consortium of Taiwan, the University of 
Wisconsin at Milwaukee, Trinity College Dublin, Lawrence Livermore National 
Laboratories, and IN2P3, France. Operations are conducted by COO, IPAC, and UW.

This work has made use of data from the European Space Agency (ESA) mission
{\it Gaia} (\url{https://www.cosmos.esa.int/gaia}), processed by the {\it Gaia}
Data Processing and Analysis Consortium (DPAC,
\url{https://www.cosmos.esa.int/web/gaia/dpac/consortium}). Funding for the DPAC
has been provided by national institutions, in particular the institutions
participating in the {\it Gaia} Multilateral Agreement. We acknowledge ESA {\it Gaia}, DPAC and the Photometric Science Alerts Team (\url{http://gsaweb.ast.cam.ac.uk/alerts}).

This publication makes use of data products from the Wide-field Infrared Survey Explorer, which is a joint project of the University of California, Los Angeles, and the Jet Propulsion Laboratory/California Institute of Technology, funded by the National Aeronautics and Space Administration. This publication also makes use of data products from NEOWISE, which is a project of the Jet Propulsion Laboratory/California Institute of Technology, funded by the Planetary Science Division of the National Aeronautics and Space Administration.

We thank the referee for comments that helped improving
the manuscript and J. Necker for sharing their catalog data.
This research is supported by the National Natural Science Foundation of China 
	(12273033) and the National SKA program of China 
	(No. 2022SKA0130101).  L.Z. acknowledges the support of the science 
	research program for graduate students of Yunnan University 
	(KC-24249083).

\end{acknowledgments}

\bibliography{tde}{}
\bibliographystyle{aasjournal}

\appendix

\restartappendixnumbering

\section{MOSFiT fitting results}
\label{sec:mf}

\begin{figure*}
    \centering
    \includegraphics[width=0.49\linewidth]{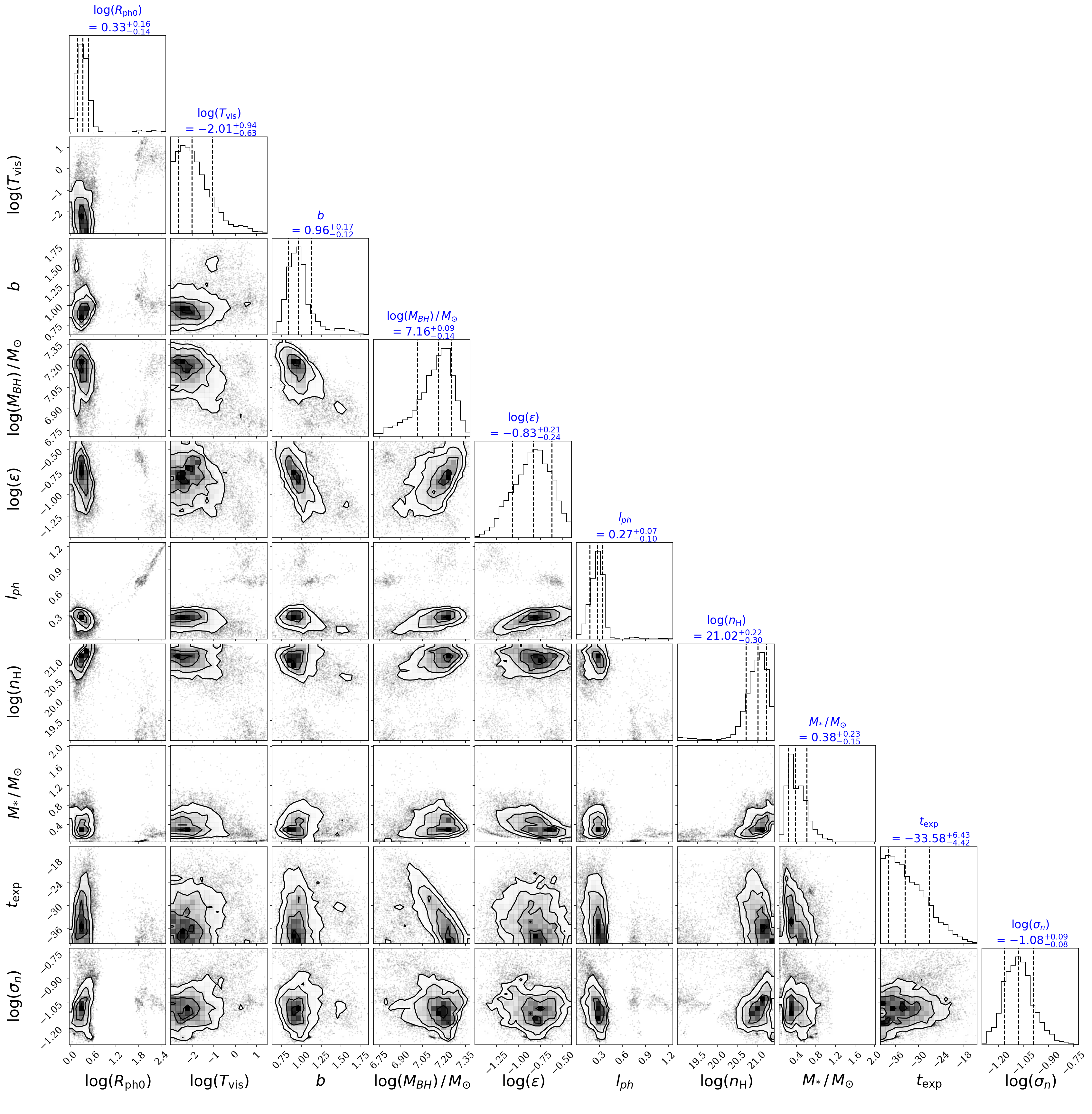}
    \includegraphics[width=0.49\linewidth]{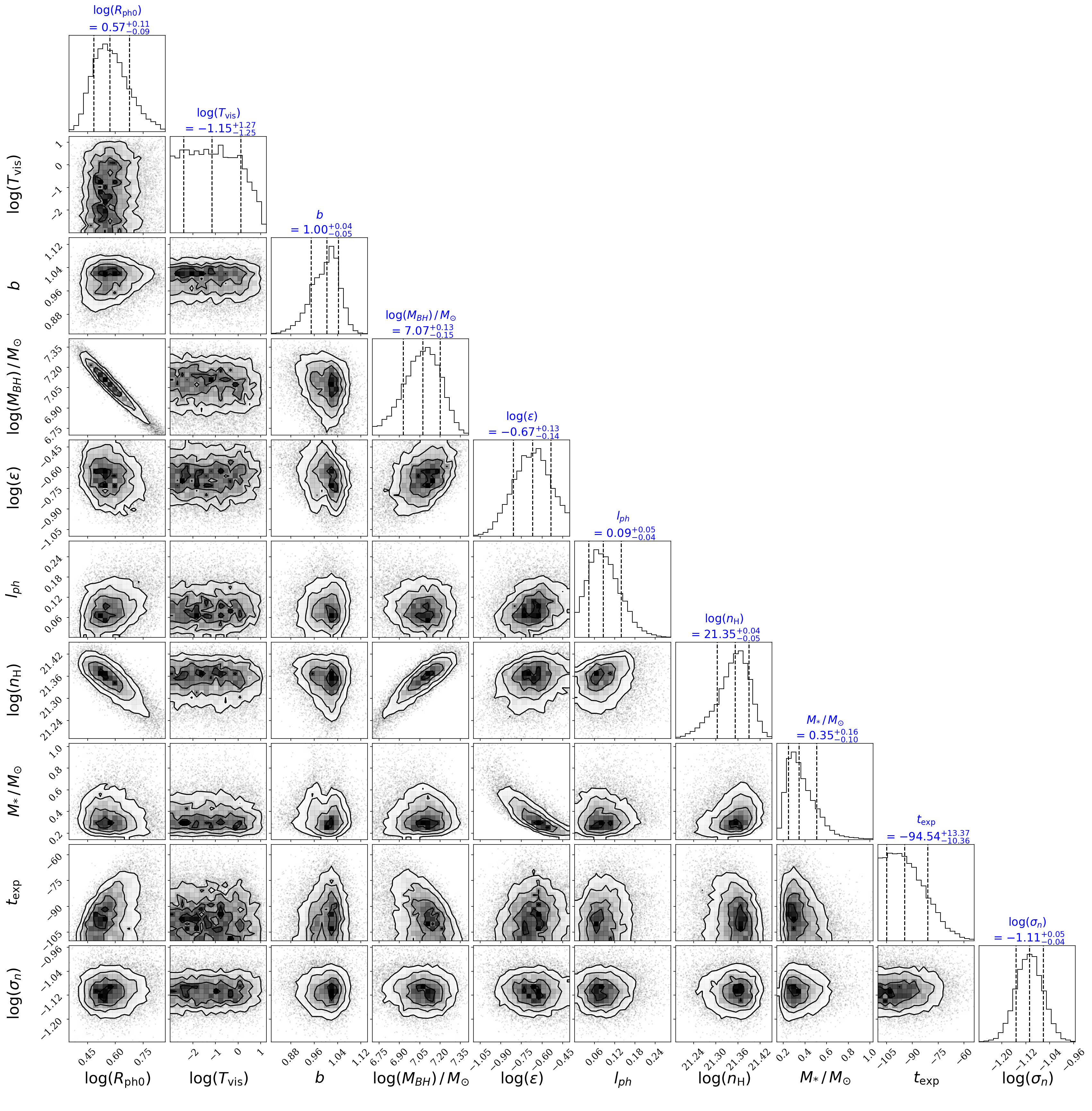}
    \caption{Posterior distributions of model parameters from the {\tt MOSFiT} fitting to flare~1 ($left$) and flare~2 ($right$) with the TDE model. The uncertainties of parameters are the 16th and 84th percentiles (1$\sigma$).}
    \label{fig:corner}
\end{figure*}

\section{PyQSOFit spectral fitting}
\label{sec:sf}

\begin{figure*}
	\centering
	\includegraphics[width=0.86\linewidth]{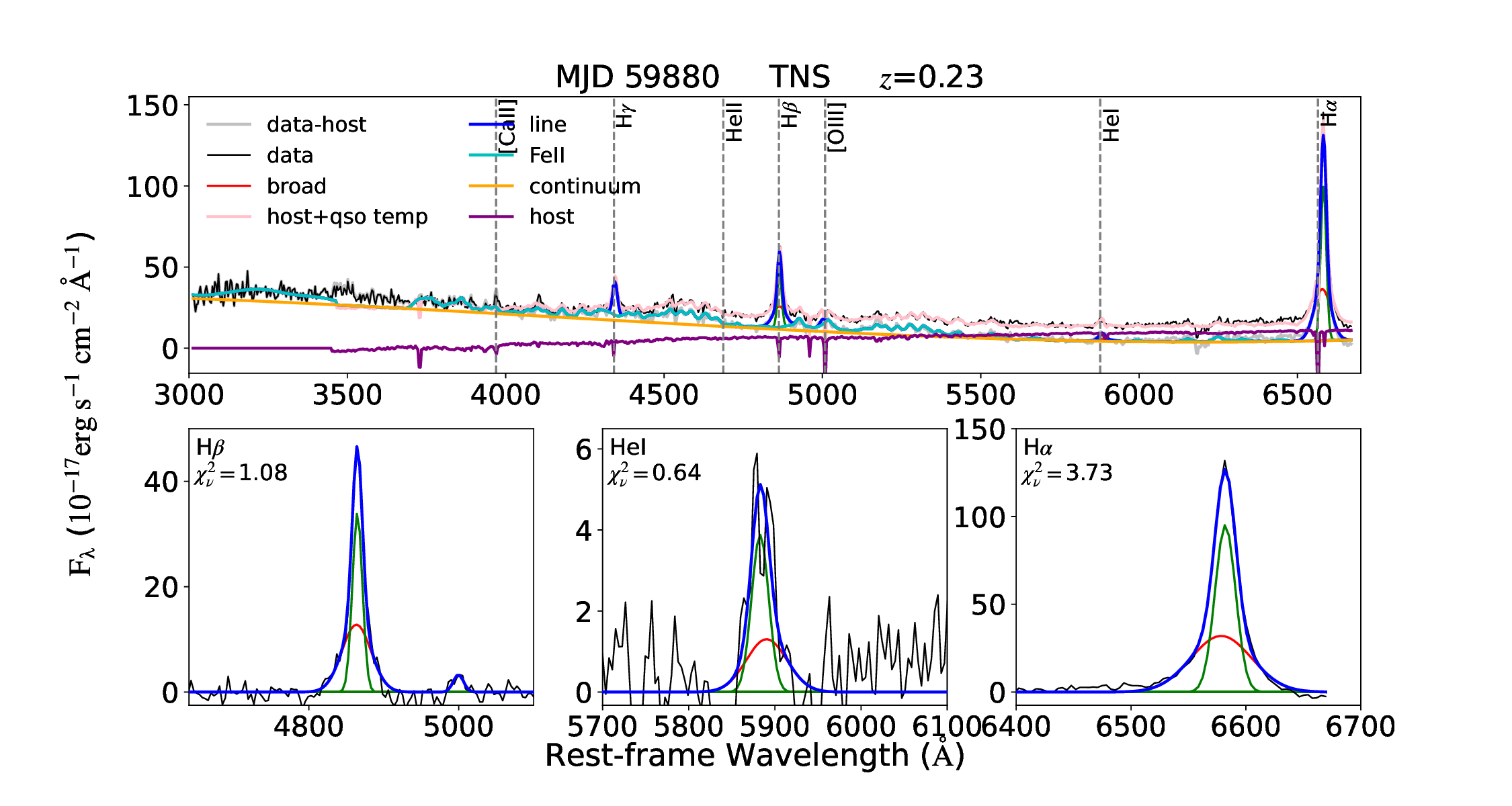}
	\includegraphics[width=0.86\linewidth]{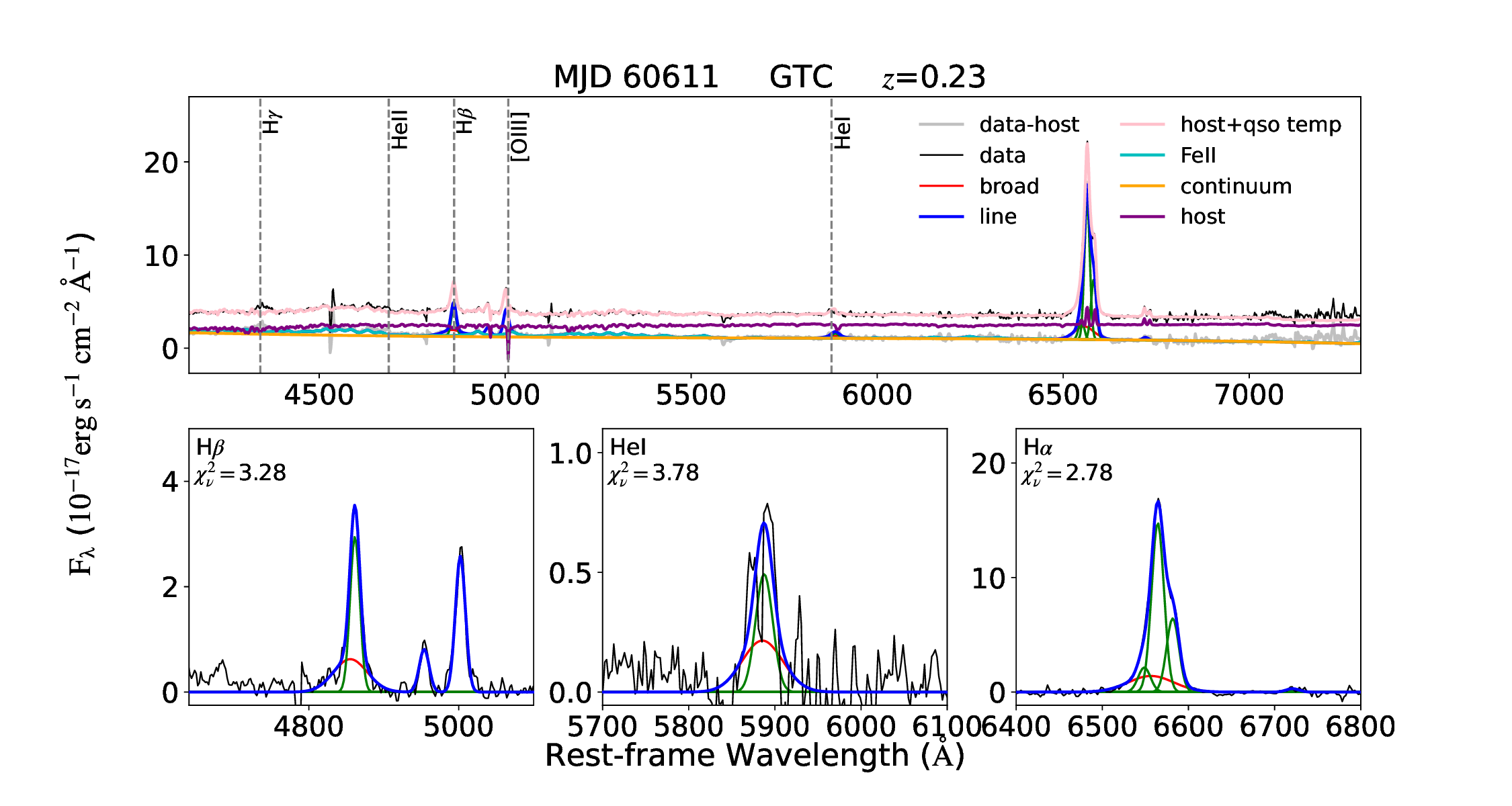}
	\caption{Spectrum fitting in PyQSOFit.}
	\label{fig:sf1}
\end{figure*}

\end{document}